\documentclass{article}
\usepackage[T1]{fontenc} 
\usepackage[utf8]{inputenc} 
\usepackage{ismir,amsmath,cite,url}
\usepackage{graphicx}
\usepackage{color}
\usepackage{listings}
\usepackage{xcolor}
\usepackage{courier}
\usepackage{lineno}
\usepackage{booktabs}
\usepackage{stfloats}
\usepackage{makecell}
\makeatletter

\newcommand{\Rmnum}[1]{\expandafter\@slowromancap\romannumeral #1@}
\makeatother

\usepackage[bookmarks=false]{hyperref}
\usepackage{xcolor}
\definecolor{BasicBlue}{RGB}{0, 0, 128}
\hypersetup{
    colorlinks,
    linkcolor=BasicBlue,
    citecolor=BasicBlue,
    urlcolor=BasicBlue
}

\title{AnimeTAB: A new guitar tablature dataset of anime and game music}

\threeauthors
  {Yuecheng Zhou} {Communication University of China }
  {Yaolong Ju} {Mcgill University}
  {Lingyun Xie} {Communication University of China }

\begin{document}
\maketitle
\definecolor{RED}{RGB}{255,0,0}
\begin{abstract}
 While guitar tablature has become a popular topic in MIR research, there exists no such a guitar tablature dataset that focuses on the soundtracks of anime and video games, which have a surprisingly broad and growing audience among the youths. In this paper, we present AnimeTAB, a fingerstyle guitar tablature dataset in MusicXML format, which provides more high-quality guitar tablature for both researchers and guitar players. AnimeTAB contains 412 full tracks and 547 clips, the latter are annotated with musical structures (intro, verse, chorus, and bridge). An accompanying analysis toolkit, TABprocessor, is included to further facilitate its use. This includes functions for melody and bassline extraction, key detection, and chord labeling, which are implemented using rule-based algorithms. 
 We evaluated each of these functions against a manually annotated ground truth. Finally, as an example, we performed a music and technique analysis of AnimeTAB using TABprocessor. Our data and code have been made publicly available for composers, performers, and music information retrieval (MIR) researchers alike.
\end{abstract}

\section{Introduction}\label{sec:introduction}
As one of the guitar score notations, tablature can be traced back in 16th century, when it is originally known as \textit{Cifra}, meaning \textit{something written in code} in Arabic \cite{alves2015history}. Cifra uses numbers on horizontal lines to represent fret positions on guitar fingerboard, and it has gradually evolved into today's guitar tablature. This simple and intuitive way of notation enables beginners to play the instrument easily, without requiring complicated music theory knowledge. Interestingly, we found an overlap between fingerstyle guitar players and Japanese anime fans\cite{japanese1,japanese2,japanese3}. While not a popular field in the world-music pantheon, Japanese pop music, anime, and video games have a surprisingly broad and growing audience among Asia youths especially young guitar players. 

\begin{figure}
    \centering
    \includegraphics[width=1.0\columnwidth]{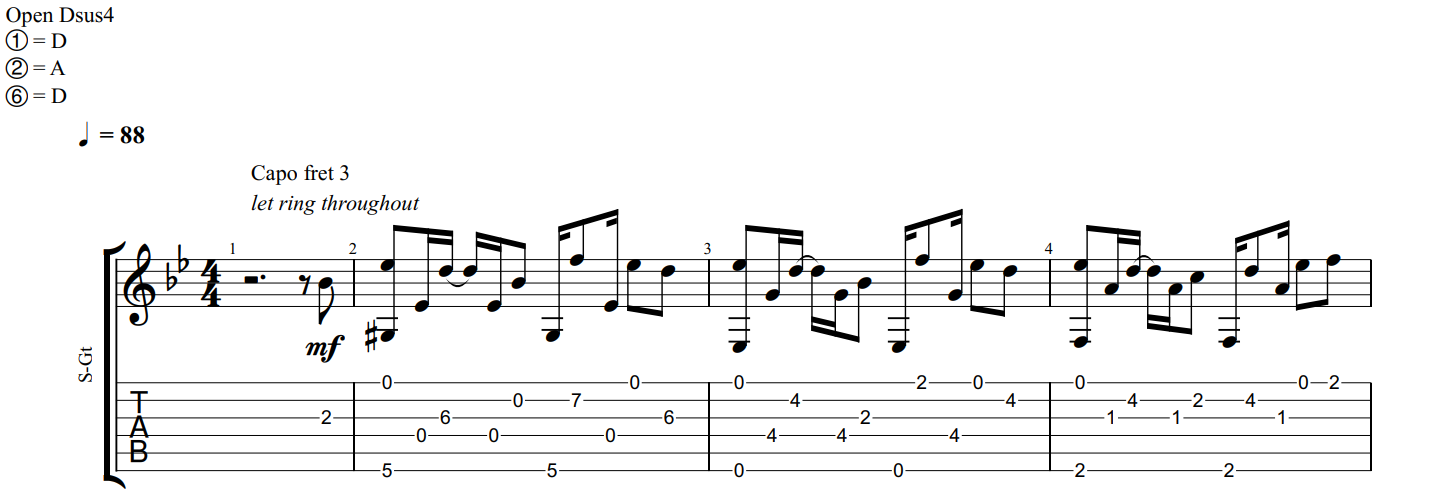}
    \caption{An example of staff and guitar tablature with special tuning and capo.}
    \label{tabfigure}
\end{figure}
In recent years, researchers have also shown interest in guitar music and tablature \cite{transchords,transrealtime, convochord,GAbased,chen2020automatic,autoleadguitar,regnier2021identification}. However, high-quality symbolic guitar tablature is not easy to obtain, especially for anime music, and this limitation has likely limited the computational study of guitar tablature to date. 
Table \ref{tab:datasets today} lists all the open-source symbolic guitar score datasets we found\cite{DADAGP, guitarset, burlet2013robotaba, chen2022towards, IDMTl}. 
Sarmento \textit{et al}. \cite{DADAGP} presented a dataset of about 26,000 band scores in GuitarPro\footnote{https://www.guitar-pro.com/} and encoded format, named DadaGP. However, the encoding tool they released is only suitable for GuitarPro 3, GuitarPro 4, and GuitarPro 5, which have long since ceased to be used (the latest version has been released to GuitarPro 8), and most tracks of DadaGP are for full bands rather than individual fingerstyle players.
Chen \textit{et al}. \cite{chen2022towards} proposed a new dataset named EGDB, which contained transcriptions of the electric guitar performance of 240 tablatures rendered with different tones.
Xi \textit{et al}. \cite{guitarset} presented a guitar music transcription dataset: GuitarSET, which including styles of rock, bossa nova and jazz. 
Burlet \textit{et al}. \cite{burlet2013robotaba} compiled two ground-truth datasets for polyphonic transcription and guitar tablature arrangement from manual transcriptions.
Kehling \textit{et al}. \cite{IDMTl} presented a novel dataset of electric guitar recordings with extensive annotation of note parameters. 
As shown above, none of them focus on fingerstyle arrangements for Japanese anime music or provide musical structure annotations. To this end, we propose a new guitar tablature dataset and an accompanying analysis toolkit. Three main contributions are presented in this paper: 
\begin{itemize}
    \item We introduce AnimeTAB, a dataset of fingerstyle guitar tablature in MusicXML format, to provide more high-quality guitar tablature for both guitar players and researchers. It contains 412 full tracks of guitar tablature from anime or video game music. 
    \item We choose 202 tablature from all the 412 tracks and segment them into 547 clips, labeling each with a start/end bar number and one of four musical structures: intro, verse, chorus, and bridge. These annotations can be used for a variety of MIR tasks, such as musical structure segmentation, music similarity, and automatic composition. 
    \item We also present TABprocessor, an analysis toolkit for AnimeTAB. This contains several rule-based algorithms to identify keys, chords, melodies, and basslines. These generated annotations are computationally inexpensive, and our experiments show that they maintain a high degree of accuracy. Additionally, we use TABprocessor to present different music statistics of AnimeTAB, including frequency distributions of keys, chords, fingering, and guitar techniques.
\end{itemize}
This paper is structured as follows.
Section 2 gives an overview of some published guitar symbolic datasets, section 3 introduces the basic information
of our AnimeTAB dataset; section 4 describes the principles and functions of TABprocessor; section 5 details the
use of TABprocessor to analyze our dataset for an example, including analysis of notes, harmonies and techniques.
Conclusions are presented in section 6. We released all the tracks, demos, and source code in [anonymized], where all the test results can also be found.

\begin{table*}[ht]\footnotesize

  \centering
  \caption{Information of open-source guitar symbolic datasets.}
    \resizebox{\textwidth}{!}{
    \begin{tabular}{l l l r l}
    \toprule[1pt]
    \textbf{Name}  & \textbf{Size} & \textbf{Type}  & \textbf{Genre} & \textbf{Format} \\
    \midrule[0.5pt]
    DadaGP\cite{DADAGP} & 26000 & symbolic & 739 different genres & GP3/GP4/GP5 \\
    GuitarSet\cite{guitarset} & 30 & symbolic/audio & rock/jazz/funk/bossa nova & JAMS \\
    ROBOTABA\cite{burlet2013robotaba} & 75 & symbolic/audio & ultimate guitar top 100 list & musicXML/MEI \\
    EGDB\cite{chen2022towards}  & 240 & symbolic/audio & solo/arpeggios & tablature and audio \\
    IDMT-SMT-GUITAR\cite{IDMTl} & 64 & symbolic/audio & 6 different genres & tablature and audio \\
    AnimeTAB(Ours)  & 412 full/547 clips & symbolic & anime & musicXML \\
    \bottomrule[1pt]
    \end{tabular}%
    }
  \label{tab:datasets today}%
\end{table*}%

\section{AnimeTAB dataset}

AnimeTAB is a structure-labeled guitar tablature dataset of anime and video game music, which can be used as symbolic data for music generation, song classification and recommendation, automatic accompaniment, phrase separation, and other MIR tasks while satisfying the daily practice needs of anime fingerstyle guitar players. The dataset consists of 412 scores and 547 labeled clips from anime episodes and video game theme songs from the 1990s to present. 
All tracks are collected from open-source scores on the Internet, most of which are written and voluntarily shared by the guitar players in anime fans community. 
Anime lovers always have an exceptional enthusiasm to share their preferences and productions, but the overall quality of their sharing are sometimes unstable.
We manually checked every score to ensure the quality and consistency, and corrected the wrong note, illegal notations, inappropriate format, and other mistakes.
Metamessages like 
Due to the requirements of different tasks, the AnimeTAB dataset is divided into full scores, score clips, and encoded scores, which will each be introduced in detail.

\subsection{Part 1:Full Scores}
The first part is a set of about 412 full scores.
These are derived from GuitarPro's native GTP format and were exported to MusicXML format using GuitarPro 7. We faithfully recorded and retained all the aspects of these tracks including special tunings, capo positions, and time signatures for special requirements in the future. Since many of the notations are specific for guitar techniques in GuitarPro software, to reproduce the complete track information, we do not recommend using other score editing software like Musescore to open them.

\subsection{Part 2:Score Clips}
Instead of entire tracks, specific music phrases like verses or chorus are more suitable in some special applications, for example, soundtracks for short videos or game scenes. The authors and another five students with background in music have selected 202 from all 412 scores, extracting specific music structures--intro, verse, bridge, and chorus--and labeled them with their structure types and starting/ending bar numbers. There are 4-16 measures in each of these slices, meta messages from the origin songs were retained. For the sake of data consistency in subsequent analysis, we set the capo\footnote{A device used on the neck of a stringed (typically fretted) instrument to transpose and shorten the playable length of the strings—hence raising the pitch} frets to zero for all clips while maintaining the fingering. More specific information is given in Table \ref{clip numbers}. Due to our careful selection, these score clips can also serve as practice tracks for guitar players.
\renewcommand{\arraystretch}{1.5}
\begin{table}[hb]
    \centering
    \resizebox{\columnwidth}{!}{
    \begin{tabular}{ l r r r}
    \toprule[1pt]
     \textbf{Structure} & \textbf{Number} & \textbf{Average bar number} &\textbf{Average note number}\\
    \midrule[0.5pt]
     Intro & 82 & 8.07 & 51.86\\
     Verse & 203 & 13.07 & 88.54\\
     Chorus & 184 & 13.01 & 86.11\\
     Bridge & 78 & 9.08 & 60.46\\
    \bottomrule[1pt]
    \end{tabular}
    }
    \caption{Information of musical structure clips.}
    \label{clip numbers}
\end{table}

\begin{figure*}[th]
    \centering
    \includegraphics[width=2\columnwidth]{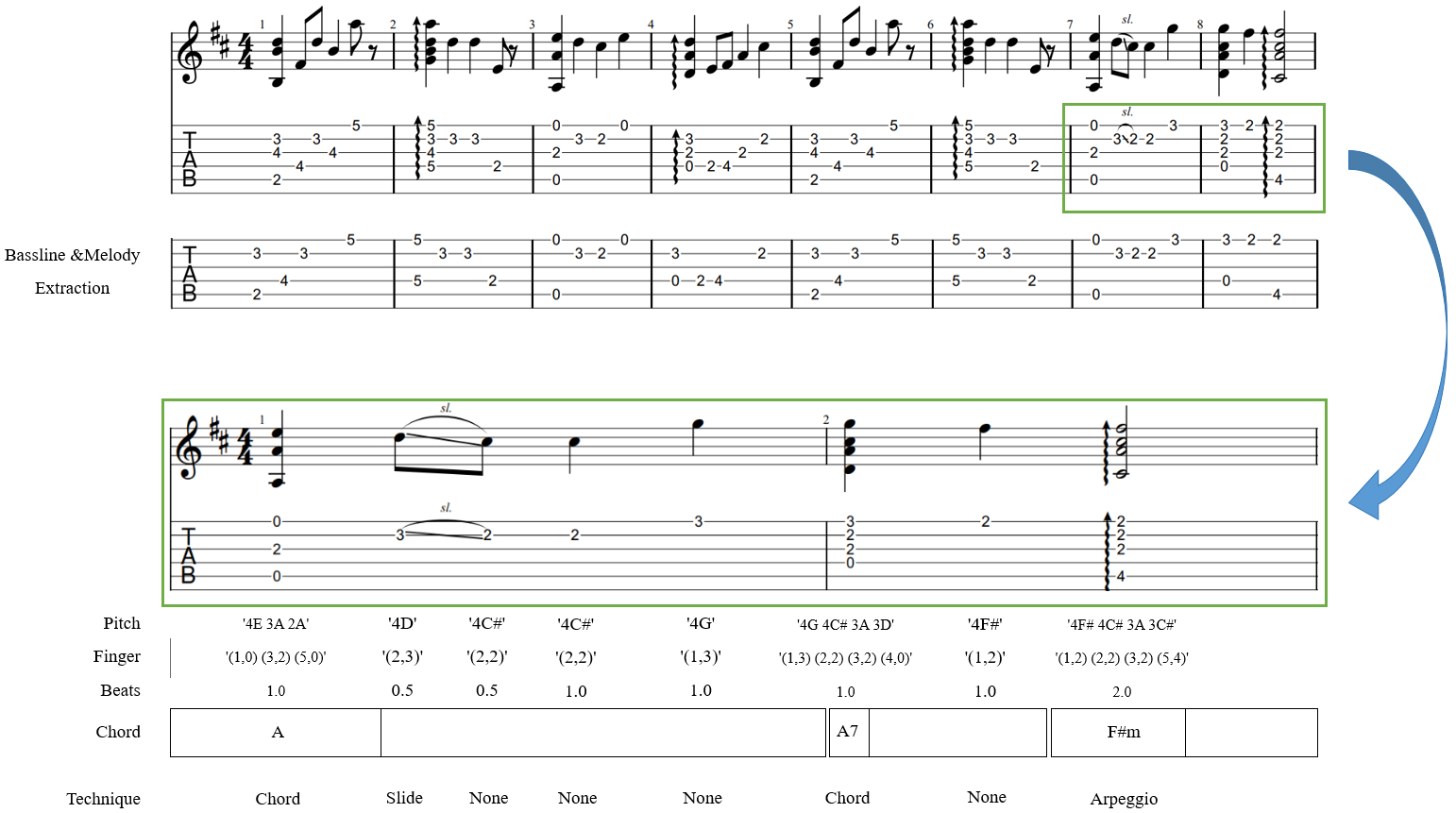}
    \caption{Bassline \& melody extraction, note encoding, chord recognition, and technique detection. Users can choose whether the note-encoding format is octave-pitch-accidental (e.g., 4C\#) or pitch-octave-accidental (e.g., C4\#). }
    \label{fig:totaltab}
\end{figure*}

\subsection{Part 3:Preprocessed Scores}
Well-known natural language processing (NLP) models used in MIR fields, such as Transformer\cite{transformer}, BERT\cite{bert},and GPT-2\cite{GPT}, can regard encoded musical tokens as actual texts for further processing. Moreover, Encoded data can take up less space and circumvent copyright issues. Precisely for these reasons, We extracted the pitch, beats, and fingering information from the 547 clips and designed this into the following format.

\begin{itemize}
    \item \textbf{Pitch.} In every note in MusicXML files, the \texttt{step}, \texttt{octave}, and \texttt{alter} nodes under the \texttt{pitch} node represent the tone name, octave, and whether there is an accidental, respectively. We encode notes into string form, and the ascending C of the fourth octave can then be encoded as "C4\#" or "4C\#", while all rests are encoded as "R". Simultaneously sounded notes are connected by spaces from high to low to form a note cluster in string form. With note encoding, users can directly put scores into the embedding layer just like actual texts in English.
    
    \item \textbf{Beats.} Note duration is recorded in MusicXML in an entirely different form than in MIDI. The number of beats of a note is determined by the ratio of its \texttt{duration} node and the bar's \texttt{division} node for a given BPM\footnote{Beats per minute}. We record this ratio as a time stamp and put it into the time list as a float32 format.
    
    \item \textbf{Finger positions.} GuitarPro 7 creates a backup for each bar with \texttt{fret} and \texttt{string} nodes under the \texttt{technical} node by default. Similar to pitch, a finger position is represented by \texttt{(string, fret)}(rests are '(R,R)'). Simultaneously sounded finger positions are connected by spaces from high to low to form a cluster of finger positions. The final result is as shown in Figure \ref{fig:totaltab}. 
    
    \item \textbf{Special tunings.} Certain strings may be tuned down a few semitones to simplify the chords and fingering in some tracks, for example in drop D\footnote{Lower the sixth string a whole tone} or open D\footnote{Adjust the six strings to D, A, D, F\#, A, D}. This is called special tuning. We record the number of semitones by which string is raised or lowered in the tuning list to avoid mismatches between the pitch and fingering. 
\end{itemize}

Each of the above text tags can be automatically converted by our toolkit-TABprocessor, and we will introduce this the next section.

\section{TABprocessor toolkit}
Currently, there are MIR toolkit like music21\cite{music21} that can extract information from musicXML files, but music21 is more oriented towards MuseScore, Finale, or Sibelius but not Guitarpro. However, due to the highly customizable character of MusicXML, it can behave completely differently when generated by different software, and we have found that a large part of scores in AnimeTAB cannot be opened by music21. That's exactly why we created our own One-click data transfer and analysis toolkit: TABprocessor.

Written in Python, TABprocessor is a lightweight MIR toolkit for information extraction, processing, and analysis functions that is specifically oriented to guitar MusicXML scores generated by GuitarPro. It is designed to allow even code beginners or musicians who do not have strong coding skills get started quickly and have fun with it. A short line of \texttt{pitchset, fingerset, timesigset, tunings = readTAB(path)} is all that is required to convert the pitches, finger positions, and note beats of an entire MusicXML score into the encoded format mentioned above. To meet the researcher's need for data analysis, We have also added other practical methods to TABprocessor for different research tasks and model frameworks, and these are introduced below.

\subsection{Bassline and Melody Notes Extraction}
Melodies and basslines are common objects in MIR research, but manual annotation of large databases is a tedious task. Thus, we provide methods for bassline and melody extraction in TABprocessor. These can be used as both an MIR analysis tool and a fingering simplification method for guitar beginners when faced with complicated scores. Due to the principle and structure of the guitar arrangement, we chose a simple but efficient rule for our extract function: 
\begin{itemize}
    \item \textbf{Bassline extraction.} When a single note or the lowest note of a note cluster is lower than the fifth string, it will be considered to be a bassline note; if fewer than two bassline notes have been extracted, the fourth-string notes will also be regarded as bassline notes to avoid a lack of low frequencies.
    \item \textbf{Melody extraction.} When a single note is on the first or second string, it will be considered a melody note. A note with a fret higher than five will be considered one string higher until the fret drops below 5. The highest note of a note cluster is considered a melodic tone regardless of its string. 
\end{itemize}
A simple demonstration is shown in Figure \ref{fig:totaltab}. Here, we chose a song with 50 bars and manually labeled the bassline and melody for testing. The results are shown in Table \ref{recall and precision}. As can be seen, this method performs well on melody notes, 
and it sacrifices accuracy for higher recall of bassline notes, of which it is better to have an excess than a shortage. Taken together, this can produce satisfactory extraction from most guitar scores. 

\renewcommand{\arraystretch}{1.5}
\begin{table}[hb]
    \centering
    \begin{tabular}{ l r r }
    \toprule[1pt]
     \textbf{Type} & \textbf{Recall} & \textbf{Precision}\\
    \midrule[0.5pt]
     Melody & 100\% & 92.42\%\\
     Bassline & 90.48\% & 65.52\%\\
    \bottomrule[1pt]
    \end{tabular}
    
    \caption{Recall rate and precision of bassline/melody extraction.}
    \label{recall and precision}
\end{table}

\subsection{Key Detection and Transfer}
The functions of key detection and transfer are provided by other similar symbolic algorithms \cite{Keyfinding1,Keyfinding2, Keyfinding3, Keyfinding4}, but in our software, we have adopted an easier and faster approach. First, the tones are differentiated by the maximum of the twelve $S_n$ values:

\begin{align}
{\rm key} = \max \left\{ \sum\limits_{n \in {K_{\rm C}}}^{} {{S_n}}, \sum\limits_{n \in {K_{{\rm C}\sharp}}}^{} {{S_n}}, \sum\limits_{n \in {K_{\rm D}}}^{} {{S_n}},\ \ \ldots, \right. \nonumber \\
\left. \sum\limits_{n \in {K_{\rm B}}}^{} {{S_n}} \right\},
\end{align}
where $S_n$ represents the total number of tones of pitch $n$, and $K_{\rm C}$ represents all in-key tones in the key of C. Major and minor are differentiated by the number of fifths and sixths of the scale. This can be useful when analyzing chord degrees or rebuilding fingerings in other keys. We created a key-labeled validation set of 45 tracks, in which our method only took about 0.06~ms for each track, while music21 took about 20~ms, and we achieved a 95.5\% correct rate in tone differentiation and 74.4\% in major/minor differentiation. We analyzed some of the error samples, and the results showed that the keys of most of the error samples were blurred and difficult to classify as strictly major or minor. This method greatly accelerates the operation speed while still being precise enough for conformist genres such as anime fingerstyle.

\begin{figure*}[th]
\includegraphics[width=2.0\columnwidth]{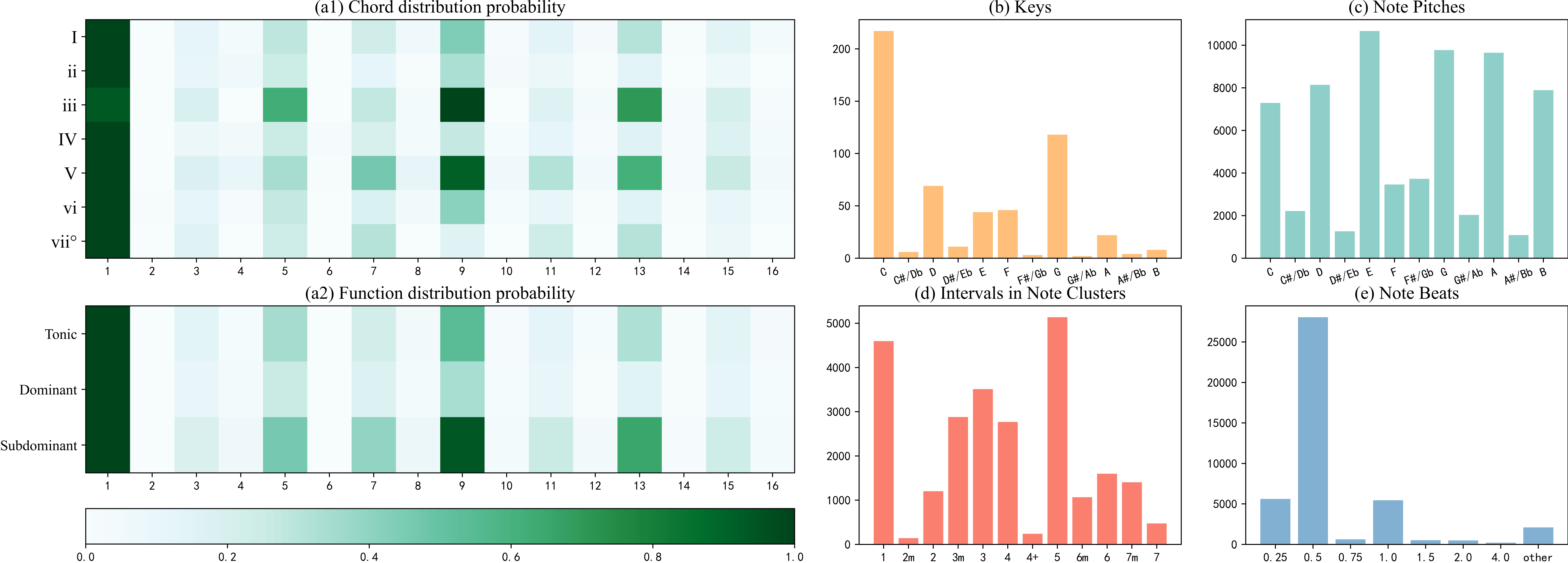}
\caption{Analysis of 560 score clips: (a)~probability distributions of chords of different step degree (upper)/function (lower) in a 4/4 bar (16th-note granularity); (b)--(e)~distributions of different keys, pitches, intervals, and special chords in the AnimeTAB dataset.}
\label{fig:bars}
\end{figure*}

\subsection{Rule-based Chord Recognition}
Chords are an important part of symbolic music. Compared with the complex spectral information that is examined for deep-learning methods using audio signals, identifying chords in symbolic music is much easier. We provide a method for detecting the type and root note of a chord by the intervals of the arranged in-chord tones. A similar function is provided by GuitarPro, but this cannot be accessed by an independent function call. Our method only considers the notes on the third, fourth, fifth, and sixth strings, because melody notes are often distributed on the first and second strings. The applicable chord types and their loop sections are listed in \tabref{tab:loop sections}. We chose 128 note clusters and removed eight chords of unsupported types for testing. During our test, it correctly identified 114 chords with an accuracy of 92.5\%. Since the \Rmnum{1}sus2 chord has the same notes as the \Rmnum{5}sus4 chord, the accuracy will increase to 95.0\% if we ignore this. Other errors occur on incomplete suspended chords, for example Asus2(no5) with [4B, 4A, 4A] will not be recognized as a chord.

\renewcommand{\arraystretch}{1.5}
\begin{table}[hb]\footnotesize
    \centering
    \begin{tabular}{ c|l l l l c l l }
    \toprule[1pt]

     \textbf{Chord}& \textbf{maj} & \textbf{min} & \textbf{maj7} & \textbf{min7} & \textbf{7} & \textbf{aug} & \textbf{dim}\\
    \midrule[0.5pt]
     Section & \makecell{43\textbf{5}\\7\textbf{5}\\4\textbf{8}}
     & \makecell{34\textbf{5}\\3\textbf{9}} 
     & \makecell{434\textbf{1}\\47\textbf{1}\\74\textbf{1}} 
     & \makecell{343\textbf{2}\\7\textbf{2}3\\73\textbf{2}}
     & \makecell{433\textbf{2}\\\textbf{2}46}
     & 444 
     & 33\textbf{6}\\
    \bottomrule[1pt]
    \end{tabular}
    \caption{Distinguishing chords by the intervals occurring. A chord's root is the pitch after the bold number (except in augmented chords). For example, a C chord (C, E, G, C) will give intervals of 4, 3, and 5 semitones, representing a major triad, and the pitch after 5 is the chord's root.}
    \label{tab:loop sections}
\end{table}

\subsection{Special Event Detection}
The guitar has a wealth of playing techniques that can greatly enrich the listening experience. We provide a function to detect chords, rests, triplets, legato lines, and techniques such as muting, artificial or natural harmonics, and glides by checking each \texttt{note} node in a bar. Users can change the dictionary to analyze other techniques if required.

\subsection{Ligato Lines and Multi-voice Processing}
In fingerstyle performance, performers always hold notes especially bass notes sustain as long as possible to gain a fuller sound effect until the next note on the same string. This results in many ligato lines between note clusters being practically useless. Thus we provide a function to clean the useless ligato lines while keeping tracks sounding the same. Also, Guitarpro allows users to edit one soundtrack in different voices so that users can write root notes, melody, and filling notes separately, but may be unreadable for other scorewriter software. we provide a method that integrate different voices or even different tracks into one. These two functions have been integrated into \texttt{readXML()}, but can be called separately.

\subsection{Conversion Function}
We provide a variety of functions to convert between the string formats, MIDI values, and guitar finger positions mentioned above. This will be helpful when faced with a mismatch between notes and fingering, such as the use of a capo or special tunings. The same MIDI value may have multiple positions on the fretboard, so this method will return multiple values from lower to higher tones.

All of the above features are designed to be as simple, intuitive, and fast as possible. We hope that our toolkit is simple and efficient enough to lower the barrier for anime and guitar enthusiasts, and that it might attract more talented people to join MIR research. For more information, please visit [anonymized]

\begin{figure*}
    \centering
\includegraphics[width=1.8\columnwidth]{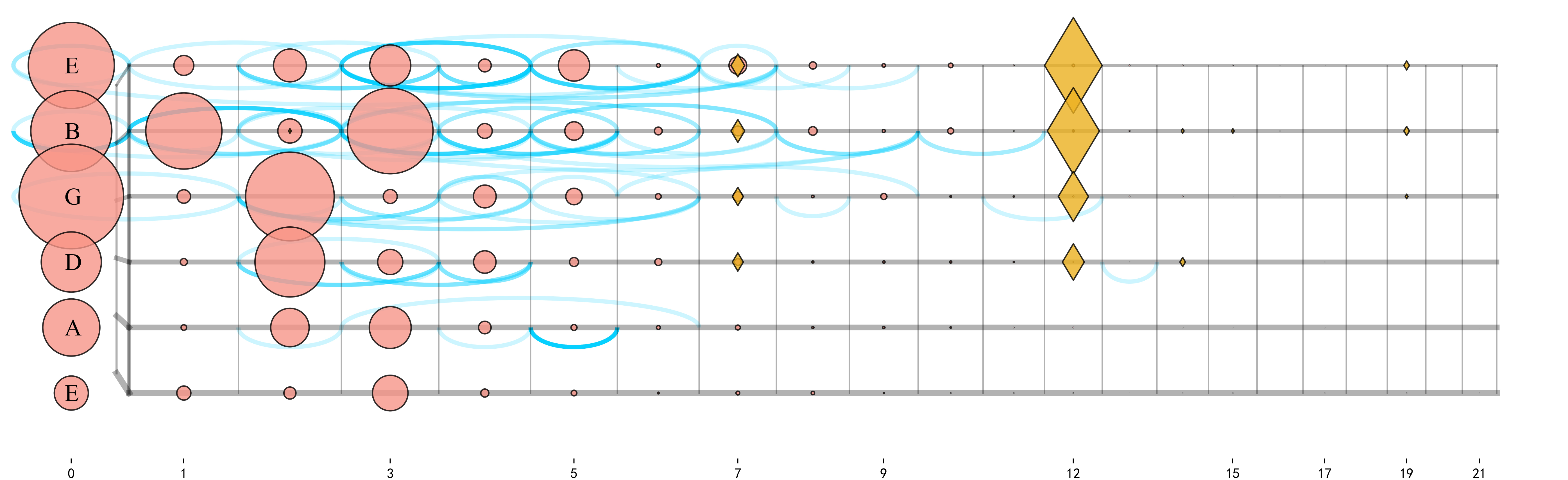}
\caption{Distribution of different guitar techniques on the fingerboard, including all notes (pink circles), natural harmonics (yellow diamonds), and slides (blue elliptical arcs).}
\label{fig:fingerboard}
\end{figure*}

\section{Analysis}
In this section, we present the use of some of the functions in the TABprocessor toolkit to analyze the 560 score clips of AnimeTAB as a demonstration. The results, figures, and code can be found in our GitHub repository.

\subsection{Note and Harmony Analysis}
\figref{fig:bars}(a1) and \figref{fig:bars}(a2) show the distribution of chords of different scale degrees and functional groups in a 4/4 bar, using our \texttt{chord\_recognize} method. The horizontal axis represents the different positions of the bar with 16th-note granularity, while the vertical axis shows distribution probabilities. This figure illustrates that notes are more likely to have whole beats such as 1, 5, 9, or 13 than half or quarter beats. \Rmnum{1} and \Rmnum{4} chords are more often used at the beginning of the bar, while \Rmnum{3} and \Rmnum{5} chords are more often used in the middle.

The keys and notes appearing in all tracks as extracted by \texttt{key\_detect} are shown in \figref{fig:bars}(b) and \figref{fig:bars}(c). As shown, the most commonly used keys in \texttt{capo} = 0 are C, G, D, and F, which are the lower degrees of the circle of fifths with the simplest fingering and few barre chords. \figref{fig:bars}(c) shows the distribution of all the notes appearing, and the C natural major scale makes up a large proportion of these. \figref{fig:bars}(d) shows the intervals in all note clusters, which are mainly harmonious intervals such as first, fifth, and fourth, and these are followed by major and minor thirds. \figref{fig:bars}(e) shows the distribution of the beats of notes, most of which are eighth notes.

\subsection{Fingerboard and Technique Analysis}
\subsubsection{Notes on the Fingerboard}
Pink circles in \figref{fig:fingerboard} show the probability of distribution of notes played at different positions on the guitar fretboard when standardized into \texttt{capo} = 0 (the first column is for open-string notes). As we can see, the second and third string are more often used than the others. The overall distribution on the fingerboard has an inverted right-angle trapezoid shape, which is related to the posture of the performer's left hand: when holding the guitar neck, the left thumb and index finger are roughly in the same fret position, while the remaining three fingers reach out from the first-string side of the neck to press the higher frets. Most of the notes are concentrated in the first four frets, since four frets with 24 pitches can completely encompass all the notes of a natural major scale while also avoiding excess finger movements.

\subsubsection{Natural Harmonics on the Fingerboard}
The yellow diamonds in \figref{fig:fingerboard} show the distribution of all the natural harmonics on the fingerboard. Obviously, the natural harmonics are distributed strictly at standing-wave node points such as half the string length (12th fret), one third of the string length (seventh or 19th fret), and a quarter of the string length (fifth fret). Higher harmonics are rarely used. More harmonic techniques can be seen to appear on the first and second strings, which we suspect has an explanation from a materials science perspective: the third, fourth, fifth, and sixth strings generally use a high-carbon steel string core wrapped in copper or alloy wire, but the first and second strings (and the third string of an electric guitar) have no wire wrapping and the string core remains exposed; this results in harmonics sounding clearer.

\subsubsection{Slides on the Fingerboard}
The blue arcs in \figref{fig:fingerboard} show all the positions at which the sliding technique appears; the lower semicircles indicate sliding from low to high, and the upper semicircles indicate sliding from high to low. The blue arcs are found mainly on the high strings and between the third and sixth frets; this indicates that most of the slide techniques are used for melodic or filler notes by the middle or ring finger. The most common distances of the slides are two or four frets, but there are a few longer slides, such as from the second to the seventh fret of the third string, and these are mostly found at a point when the melody changes quickly between verse and chorus.

\section{prospective applications}

\subsection{Use Case: Music Generation}
Music generation tasks especially those using deep-learning methods have attracted widespread attentions of countless MIR researchers in recent years. From the basic Perceptron to the state-of-the-art NLP models, the utilization of deep-learning in music generation tasks has become a trend, and has been proved success in many previous study. 
Huang \textit{et al}. \cite{musictransformer} demonstrated a Transformer model with a modified relative attention mechanism to generate minute-long compositions on a given motif. 
Yang \textit{et al}. \cite{midinet} pioneered the application of convolutional neural networks (CNNs) to symbolic-domain music generation task, making it possible to generate melodies both from a chord sequence from scratch or from previous notes by conditioning. Obviously, new high-quality data is never too much for deep-learning models.

\subsection{Use Case: Music Structure Analysis}
Music elements can range from local/short-term levels (like motif) to long-term levels (like segments). Longer temporal scales like verse or chorus may contain underlying messages that could be useful in various tasks. There have been some muisc audio segmentation tasks like \cite{audiosegment1, audiosegment2, audiosegment3}. The performance these segmentation algorithms has been shown to be dependent on the audio features chosen to represent the audio, but annotated audio datasets are time consuming and occupies more space than symbolic datasets. In \cite{melodysegment}, Guan \textit{et al}. 
deployed various neural network architectures on symbolic representation of music to detect different segments. Experiment results show that the CNN-CRF architecture performs the best. Sawada \textit{et al}. \cite{melodysegment2} presented an unsupervised note sequences segmentation method using a language model based on a non-parametric Bayesian model. Treating music as a language, the model can conduct melody sequences without annotations. Our dataset can be used as a labeled ground truth in future similar researches.

\subsection{Use Case: Humanities and Multimodal Research of ACG Culture}
Cultural diversity is the source of creativity for MIR community. Despite being considered as a niche area, but with the fast development of media and information technology, the anime, comic and game , a.k.a ACG culture, is becoming increasingly popular during the past decades. In \cite{AnimeInfluenceChina}, by content analysis and comparing, Lu \textit{et al}. explain the impact by Japanese anime on Chinese youth, and the influence of new media and technologies on its spread; In \cite{AnimeInfluenceMala}, Yamato \textit{et al}. have analyzed the data from in-depth interviews of nine Malaysians that have been involved in ACG music and other events for more than five years.
Follow-up cultural communication researchers can deploy Multimodal analysis based on our data and metamessage.
We hope through our tireless efforts, this minority culture can be passed on for one generation to another.

\section{Conclusion}
In this paper, we present AnimeTAB, an anime guitar song dataset based on the MusicXML and GTP formats. This contains: (1)412 complete songs; (2)547 score clips from the 110 best-quality tracks, each of which is labeled with a start/end bar number and one of four musical structures (verse, chorus, bridge, or intro); (3)~encoded notes. This dataset fills the gap relating to the lack of structure-labeled anime guitar tablature dataset in the MIR field. We also present TABprocessor, our data-encoding toolkit for guitar MusicXML files. This includes practical functions such as data encoding, key and chord detection, and root and melody extraction. We have tested the performance of these functions and presented a sample analysis. In future studies, we intend that we will gradually improve the user guide and add more functions to this system while also updating the dataset.
\bibliography{ref}
\end{document}